\documentclass[twocolumn,pra,twocolumn,superscriptaddress]{revtex4}
\usepackage{color}
\usepackage{amsmath}
\usepackage{amssymb}
\usepackage{graphicx}
\usepackage{tikz}

\usepackage{float}
\usepackage{bbm}

\usepackage{epsfig} 
\usepackage{verbatim}
\usepackage{array}
\usepackage{setspace}

\usepackage[unicode=true,
 bookmarks=true,bookmarksnumbered=false,bookmarksopen=false,
 breaklinks=false,pdfborder={0 0 1},backref=false,colorlinks=true]
 {hyperref}
\hypersetup{
 linkcolor=magenta, urlcolor=blue, citecolor=blue, pdfstartview={FitH}, hyperfootnotes=false, unicode=true}
 
\usepackage{color}

\makeatletter
\@ifundefined{textcolor}{}
{%
 \definecolor{BLACK}{gray}{0}
 \definecolor{WHITE}{gray}{1}
 \definecolor{RED}{rgb}{1,0,0}
 \definecolor{GREEN}{rgb}{0,1,0}
 \definecolor{BLUE}{rgb}{0,0,1}
 \definecolor{CYAN}{cmyk}{1,0,0,0}
 \definecolor{MAGENTA}{cmyk}{0,1,0,0}
 \definecolor{YELLOW}{cmyk}{0,0,1,0}
}

\usepackage{amsfonts}\usepackage{tabularx}\usepackage{dcolumn}\usepackage{bm}\usepackage{graphicx}\usepackage{epstopdf}

\hypersetup{urlcolor=blue}

\makeatother

\newcolumntype{C}[1]{>{\centering\arraybackslash$}p{#1}<{$}}

\usepackage{algorithm}
\usepackage{algorithmic}

\begin{document}

\widetext

\title{Automatic spin-chain learning to explore the quantum speed limit}

\author{Xiao-Ming Zhang}
\thanks{These authors contributed equally}
\affiliation{Department of Physics, City University of Hong Kong, Tat Chee Avenue, Kowloon, Hong Kong SAR, China, and City University of Hong Kong Shenzhen Research Institute, Shenzhen, Guangdong 518057, China}
\author{Zi-Wei Cui}
\thanks{These authors contributed equally}
\affiliation{Institute for Quantum Science and Engineering and Department of Physics,
Southern University of Science and Technology, Shenzhen 518055, China}
\author{Xin Wang}
\email{x.wang@cityu.edu.hk}
\affiliation{Department of Physics, City University of Hong Kong, Tat Chee Avenue, Kowloon, Hong Kong SAR, China, and City University of Hong Kong Shenzhen Research Institute, Shenzhen, Guangdong 518057, China}
\author{Man-Hong Yung}
\email{yung@sustc.edu.cn}
\affiliation{Institute for Quantum Science and Engineering and Department of Physics,
Southern University of Science and Technology, Shenzhen 518055, China}
\affiliation{Shenzhen Key Laboratory of Quantum Science and Engineering, Shenzhen, 518055, China}

\date{\today}
\begin{abstract}
One of the ambitious goals of artificial intelligence is to build a machine that outperforms human intelligence, even if limited knowledge and data are provided. Reinforcement Learning (RL) provides one such possibility to reach this goal. In this work, we consider a specific task from quantum physics, i.e.~quantum state transfer in a one-dimensional spin chain. The mission for the machine is to find transfer schemes with fastest speeds while maintaining high transfer fidelities. The first scenario we consider is when the Hamiltonian is time-independent. We update the coupling strength by minimizing a loss function dependent on both the fidelity and the speed. Compared with a scheme proven to be at the quantum speed limit for the perfect state transfer, the scheme provided by RL is faster while maintaining the infidelity below $5\times 10^{-4}$. In the second scenario where a time-dependent external field is introduced, we convert the state transfer process into a Markov decision process that can be understood by the machine. We solve it with the deep Q-learning algorithm. After training, the machine successfully finds transfer schemes with high fidelities and speeds, which are faster than previously known ones. These results show that Reinforcement Learning can be a powerful tool for quantum control problems. 
\end{abstract}
\maketitle

\section{ Introduction}
The ability to shuttle information efficiently and accurately between quantum systems is key to the scalability of quantum computation and simulation \cite{Yung2012c,Kassal2010}. 
Quantum state transfer \cite{nikolopoulos.14} can be applied in quantum computation \cite{Yung2006a} and is essential for a broad range of experimental platforms including trapped ions \cite{Richerme.14,Jurcevic.14,Yung2014,Shen2015a,Zhang2015a,Zhang2016a}, cold atoms \cite{Barredo.15,Schempp.15}, quantum dots \cite{Divincenzo.2000,nikolopoulos.04, nikolopoulos042, yang.10spin,Veldhorst.15}, superconducting \cite{Hu2017} and donor qubits \cite{Mohiyaddin.16}, NMR~\cite{Li2011,Zhang2012c,Li2016a}, and photonic systems~\cite{Xu2012,bellec.12, perez.13, Peruzzo2013,nikolopoulos.15,Xu2013, Chapman.16b} etc.
Among various physical systems and geometries \cite{Yung2003a,Osborne.04,Christandl.04,Bayat.15,Korzekwa.14,maring.17,Chapman.16b,Baksic.16,Vermersch.17,Agundez.17, kostak.07, brougham.09, brougham.11, nikolopoulos.12}, the transfer of a quantum state through a one-dimensional spin chain~\cite{Bose2003,Yung.05} is among the most studied. While extensive researches have been dedicated to improve the fidelity of the transfer, including proposals accomplishing perfect transfer when certain conditions are met \cite{Yung.05,kay.06,Franco.08,08.Balachandran,Vinet.12}, relatively less attention has been paid to the efficiency, i.e.~the speed of the state transfer. 

The Heisenberg uncertainty principle imposes an upper bound of the time required by transferring a spin on an $N$-qubit chain via sequential swap operations as $\pi(N-1)/(2J)$ \cite{Deffner.17}, where $J$ is the coupling strength between neighboring spins. For a sufficiently long chain with more flexible controls,  this value has been further refined as the maximum spin-wave speed corresponding to a transfer time equaling $(N-1)/(2J)$ \cite{Ashhab.12}.
 Nevertheless, the actual Quantum Speed Limits (QSLs) for individual systems vary with the types of spin coupling, available controls, etc. The QSL for perfect spin transfer along a chain with time-independent controls has been found to occur \cite{Yung.06} when the exchange coupling is modulated as set forth by \cite{Christandl.04}, but this limit lies far below the value provided by \cite{Ashhab.12}.  
 
 The transfer speed may be further enhanced when dynamical control of the system is allowed and minimal sacrifice of the fidelity is tolerated. In a previous work \cite{Caneva.09}, a control protocol has been found numerically using the Krotov method \cite{Krotov.96} that is faster than previously known protocols while the infidelity is maintained below $10^{-4}$. Further enhancement of the quantum transfer speed remains an open question, which possibly demands development of new methodologies.

In this work, we aim at finding protocols for optimizing the quantum speed limit through machine learning. Overall, the technique of machine learning has been applied to many problems related to quantum physics \cite{Deng.17(2), Zhang.17, Deans.18, August.17, Carrasquilla.17, Wang.16, Stoudenmire.16, Torlai.16, Torlai.17, Gao, Torlai.17(2),Han.17, Han.17a, Ma.17, Lu.17, Zhang.17(2), Yang2017, Liu.17}. It can be broadly divided into two categories depending on whether extensive inputs are required. In supervised and unsupervised learning, a large amount of input data is supplied to the machine, which then finds underlying links within the data and makes predictions. On the other hand, Reinforcement Learning (RL) requires minimal human input, and an agent tries to maximize certain reward under a set of rules prescribed according to specific problems that we desire to solve. The idea of RL has been successfully applied to many problems including quantum state preparation \cite{Chen.14,Bukov.17} and tomography \cite{Mahler.13, Ferrie.14,Chapman.16}, finding the ground states \cite{Carleo.17, Deng.17}, quantum gate engineering \cite{Banchi.16}, Monte Carlo method \cite{Liu.17a, Liu.17b} and experiment design \cite{Melnikov.18}. An enhanced version of RL, termed as deep RL (to be detailed later in Sec.~\ref{sec:deep}), has recently been demonstrated to achieve super-human performances in several complex games which are long deemed difficult \cite{Mnih.15,Silver.16,Silver.17}. Meanwhile, the problem of quantum state transfer may be converted into a game, and it has been demonstrated that even human players can outperform the Krotov algorithm for certain problems allowing minimal errors from the players \cite{Sorensen.16}. These and other related researches inspire us to apply the  RL methodology to the problem of quantum state transfer which may enhance the transfer speed while keeping the infidelities to a minimum.

In this paper, we apply RL to explore the QSL of transferring a spin state from one end of a spin chain to the other.  We consider both scenarios where the model Hamiltonian may be time-independent, or time-dependent. 

In the time-independent scenario,  we apply a self-learning algorithm which minimizes a cost function related to both the fidelity and the speed of the transfer by properly modulating the coupling between the spins. We found that for short chains (less than or equal to ten spins), our numerical results always agree exactly with the standard ones~\cite{Christandl.04}. However, for longer chains (more than 10 spins), our results deviate from the standard engineering \cite{Christandl.04}. Albeit the state transfer is no longer perfect, the transfer becomes faster (by about $10\%$ for a 150-spin chain) at the cost of minimal imperfection in the fidelity (i.e. infidelity$<5\times10^{-4}$).

In the scenario for which the Hamiltonian is time-dependent, we consider dynamical control of external magnetic fields, while fixing all coupling within the chain at a constant value (i.e.~a uniform chain with the time-dependent external magnetic field). We allow the magnetic fields at each site to switch on and off in the course of operation. The problem can then be converted into a so-called Markov Decision Process (MDP) problem \cite{Sutton} suitable for deep RL. 

We have studied the Heisenberg chains with up to 11 spins and have found control schemes that lead to a faster state transfer than known results via the Krotov method. In this case, our infidelity is maintained below 5\% and only controls on the first and last few spins are required. Our results therefore suggest that there can be further improvements on the best known results on spin transfer, provided that more tolerance on the infidelity is allowed. We then discuss how our method can be optimized to achieve even higher transfer fidelities and be  generalized to study problems with longer spin chains.

The remainder of the paper is organized as follows. In Sec.~\ref{sec:model} we present the model and methods used in this work, including the time-independent case (Sec.~\ref{sec:ind}) and time-dependent case (Sec.~\ref{sec:dep}). We then present results in Sec.~\ref{sec:res}, and conclude in Sec.~\ref{sec:con}.

\section{Model and Methods}
\label{sec:model}
\subsection{Time-independent case}
\label{sec:ind}
\subsubsection{Model}
We consider a spin-$1/2$ chain with $XY$ coupling. The Hamiltonian of an $N$-spin chain is (we set $\hbar=1$ in this work): 
\begin{equation}
H(\bm{J}^N)=\frac{1}{2}\sum_{n=1}^{N-1} J^{N}_{n,n+1}(\sigma^{x}_{n}\sigma^{x}_{n+1}+\sigma^{y}_{n}\sigma^{y}_{n+1}),\label{eq:tindepHam}
\end{equation}
where $\sigma_{n}^{x},\sigma_{n}^{y}$ are Pauli $x$, $y$ matrices acting on the $n^{\rm th}$ spin,
\begin{equation}
\bm{J}^N=\left[J^{N}_{1,2},J^{N}_{2,3},\ldots,J^{N}_{N-1,N} \right]^{ T} \ ,
\end{equation}
and $J^{N}_{n,n+1}$ is the coupling strength between the $n^{\rm th}$ and $(n+1)^{\rm th}$ spins to be optimized. 

The $z$-component of the total angular momentum is conserved; we therefore restrict ourselves to states with only one spin up and all other spins in the down state. We denote the state with the $n^{\rm th}$ spin being up as  $|\textbf{n}\rangle$. 
 Under the basis $\{|\textbf{1}\rangle , |\textbf{2}\rangle \ldots  |\textbf{N}\rangle \}$, the Hamiltonian can be written in a matrix form as: 
 \begin{eqnarray}
H(\bm{J}^N)=\left(\begin{array}{cccccc}
0 & J^{N}_{1,2} & 0 &\cdots& 0\\
J^{N}_{1,2} &0 & J^{N}_{2,3}&\cdots&0\\
 0 & J^{N}_{2,3} & 0&\cdots&0 \\
 \vdots &\vdots & \vdots&\ddots &J^{N}_{N-1,N}\\
 0 & 0 & 0&J^{N}_{N-1,N}&0 \\ 
\end{array}\right).\label{eq:tindepHammat}
\end{eqnarray}
We also define $|\textbf{0}\rangle$ to be the state with all spins down.

The problem of the quantum state transfer is stated as follows. The system state is initialized as  
\begin{equation}
\alpha|\textbf{0}\rangle+\beta|\textbf{1}\rangle \ ,
\end{equation}
and is then allowed to evolve for a time duration $\tau$ under the Hamiltonian [Eq.~\eqref{eq:tindepHammat}]. The evolution operator is then 
\begin{equation}
U(\bm{J}^N, \tau)=e^{-iH(\bm{J}^N)\tau} \ .
\end{equation}

Since $U(\bm{J}^N, \tau)|\textbf{0}\rangle\equiv|\textbf{0}\rangle$, our problem is then finding a unitary relating $|\textbf{1}\rangle$ and  $|\textbf{N}\rangle$ (up to a phase factor) i.e.
\begin{equation}
U(\bm{J}^N, \tau)|\textbf{1}\rangle=e^{-i\phi}|\textbf{N}\rangle.
\end{equation}         
The phase $\phi$ is unimportant for our purpose which can be eliminated with a single qubit operation. 

The fidelity of the transfer is defined as 
\begin{equation}
F(\bm{J}^N, \tau)=|\langle \textbf{N}|U(\bm{J}^N, \tau)|\textbf{1}\rangle |^{2} \ .
\end{equation}
Our goal is to tailor the couplings $\bm{J}^N$ so that the transfer can be completed with the shortest possible $\tau$ while maintaining the fidelity $F(\bm{J}^N, \tau)$ as close to 1 as possible. 

In order to compare different transfer schemes, it is useful to introduce a quantity termed as the ``efficiency'',    $J_{\rm max}\tau$, where $J_{\rm max}=\max(\bm{J}^N)$ is the maximal coupling strength in the chain.

\subsubsection{Learning algorithm}
If the efficiency $J_{\max}\tau$ is fixed, the total transfer time $\tau$ scales inversely with the maximal coupling strength $J_{\max}$. For the convenience, we fix the total transfer time $\tau=1$ as our time unit, so that a ``faster'' transfer scheme refers to a smaller $J_{\max}$. And we denote $F(\bm{J}^N, 1)\equiv F(\bm{J}^N)$. 

On the other hand, the loss function used in RL is defined as:
 \begin{eqnarray}
L(\bm{J}^N)\equiv1-F(\bm{J}^N)+\lambda J_{\max}.
\end{eqnarray}
Since allowing for a lower fidelity always improves the efficiency so that the two are in competition, we introduce a relaxation parameter $\lambda$  controlling the weights between the fidelity and efficiency. 

Our learning algorithm for an $N$-spin chain is as follows.
\begin{itemize}
\item[{\bf Step 1:}] Randomly pick a site $m$ $(1\leq m<N)$, and generate two perturbed coupling $\bm{J}^{N+}$ and $\bm{J}^{N-}$ with 
\begin{equation}
J^{N\pm}_{n,n+1}=J^{N}_{n,n+1}\pm\beta\delta_{n,m} \ ,
\end{equation}
where $\beta$ is the perturbation strength.
\item[{\bf Step 2:}] Calculate the gradient for the loss function on the perturbed site: 
\begin{eqnarray}
g_m=\frac{  L(\bm{J}^{N+})-L(\bm{J}^{N-}) }{2\beta}. \label{eq:grad}
\end{eqnarray}
\item[{\bf Step 3:}]Update the coupling strength on the perturbed site according to
\begin{equation}
J^{N}_{m,m+1}\leftarrow J^{N}_{m,m+1}+ \alpha g_{m} \ ,
\end{equation}
where $\alpha$ is the learning rate. 
\item[{\bf Step 4:}] Repeat steps 1-3 until
\begin{equation}
1-F(\bm{J}^N)<\xi \ ,
\end{equation}
(where $\xi$ is the precision desired) is satisfied or the maximum number of iterations are reached.
\end{itemize}

We suppose that the strengths of the couplings that sustain the spin transfer would not change much if one adds or removes one spin to or from the chain. This enables us to perform the learning algorithm recursively. Suppose the optimized results have been obtained for an $(N-1)$-spin chain, the initial guess of $\bm{J}^N$ for the $N$-spin chain can be set as: 
\begin{eqnarray}
J^{N}_{n,n+1}= \left\{
\begin{array}{rcl}
J^{N-1}_{n,n+1}      &    & n\leq\frac{ N}{ 2}, \\
J^{N-1}_{n-1,n}    &  & n> \frac{N}{2}.   \label{eq:rec}\\  
\end{array} \right. 
\end{eqnarray}
A full description of the learning process is given in Algorithm~\ref{alg1}.

\begin{algorithm} 
\caption{}  
\label{alg1}  
\begin{algorithmic}

\STATE Set initial guess of $\bm{J}^{3}$ randomly

\FOR{spin number $N=3$, $N_{\rm max}$} 
\STATE

\quad\textbf{for}  iteration $k=1$, $K_{\rm max}$, \textbf{do}

\quad \quad Randomly choose a spin $m$

\quad \quad Calculate the gradient $g_{m}$ from Eq.~\eqref{eq:grad}

\quad \quad Update the coupling $J_{m,m+1}^{N}\leftarrow J_{m,m+1}^{N}-\alpha g_{m}$ 

\quad \quad \textbf{Break} if $1-F(\bm{J}^N,\tau)<\xi$

 \quad \textbf{end for} 
 
 \quad Set initial guess of $\bm{J}^{N+1}$ using Eq.~\eqref{eq:rec}
 
\ENDFOR 

\end{algorithmic} 
\end{algorithm} 

\begin{figure*}
\includegraphics[scale=0.6]{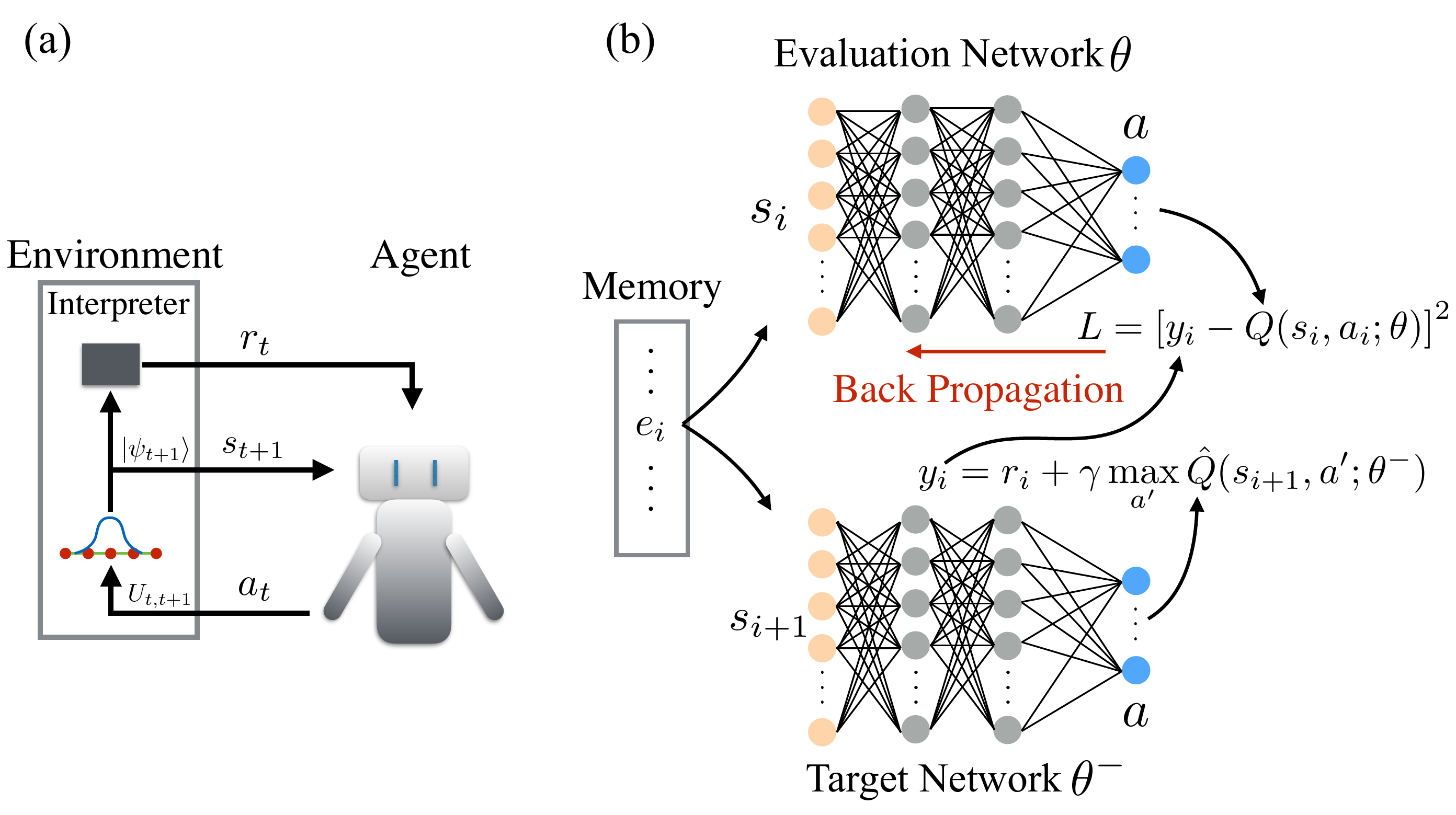}
\caption{\label{fig:epsart} Sketch of the deep Reinforcement Learning. (a) The agent chooses an action $a_{t}$ at each step. A corresponding evolution operator $U_{t}$ is applied to the current wave function. The updated wave function $|\psi_{t+1}\rangle$ generates the next state $s_{t+1}=f(|\psi_{t+1}\rangle)$ for the agent. The interpreter will output reward $r_{t}$ according $s_{t}$. The experience $(s_{t}, a_{t},r_{t},s_{t+1})$ as a whole is stored in the memory $R$ for future updating of the neural network. (b) Neural network updating process. At each step, a minibatch of experiences $(s_{i},r_{i},a_{i},s_{i+1})$ are retrieved. $s_{i+1}$ is input to the target network, which outputs  $\hat{Q}(s,a;\theta^{-})$. $s_{i}$ is input to the evaluation network, which outputs $Q(s,a;\theta)$. Then, back-propagation is performed for $\theta$ according to the loss function defined in Eq.~\eqref{eq:L}. Here we used the Rectifier Linear Unit ($\mathrm{max}(0,x)$) as the activation function for both neural networks.}
\label{fig:sketch}
\end{figure*}

\subsection{Time-dependent case}
\label{sec:dep}
\subsubsection{Model}
In this part, we fix the coupling strength along the chain to be uniform (i.e.~$J_{n,n+1}^{N}=J$ for all $n$), and allow time-dependent control via external magnetic fields ($B_n(t)$ for the $n^{\rm th}$ spin). The Hamiltonian is then given by
\begin{equation}
H^{\rm dyn}(t)=\frac{J}{2}\sum_{n=1}^{N-1} (\sigma^{x}_{n}\sigma^{x}_{n+1}+\sigma^{y}_{n}\sigma^{y}_{n+1}) +\sum_{n=1}^{N}B_n(t)\sigma^{z}_n. \label{eq:timedpHam}
\end{equation}
We note that contrary to Sec.~\ref{sec:ind} where we have fixed the total time $\tau=1$ while varying the coupling, in this section we fix $J$ in all cases and allow $\tau$ to vary. The total transfer time $\tau$ is discretized into $N^\tau$ slices with step size $dt$ and we denote each slice by $t$ that takes integer values $0,1,\ldots N^\tau$. $t=0$ therefore denotes the starting time and $\tau=N^\tau dt$. We also denote the wave function at time slice $t$ as $|\psi_t\rangle$.

The system is initialized into $|\psi_{0}\rangle=|\textbf{1}\rangle$. The evolution due to Eq.~\eqref{eq:timedpHam} is
\begin{equation}
 U_{t,t+1}=e^{-iH^{\rm dyn}(t)dt}.
\end{equation}
The state is updated from each time step, 
\begin{equation}
|\psi_{t+1}\rangle=U_{t,t+1}|\psi_{t}\rangle.
\end{equation}
Our goal is to minimize $N^\tau$ (i.e.~$\tau$) while maintaining the final fidelity $F=|\langle \mathbf{N}|\psi_{\tau}\rangle |^{2}$ as close to 1 as possible.

In \cite{Caneva.09} and \cite{Murphy.10}, those authors have assumed a parabolic shape of the magnetic field, i.e.~$B_n(t)=C(t)\left[x_n-d(t)\right]^2$, where $x_n$ is the location of the $n^{\rm th}$ spin, and $d(t)$ denotes where the parabolic potential is centered (which changes over time). The Krotov algorithm is then used to optimize $C(t)$ and $d(t)$ and the results show that the state transfer can be performed in a time that is merely a few percent longer than the one given by the maximum spin-wave speed \cite{Ashhab.12}: 
\begin{equation}
T_{\rm QSL}=\frac{N-1}{2J} ,
\end{equation}
which can be regarded as the upper bound of the time spent in the quantum state transfer.

In this work, we adapt a  different setup. For each spin, the magnetic field can only be switched ``on'' or ``off''. In other words, at each time step $t$, the magnetic field for spin $m$ can only take two values: 
\begin{eqnarray}
&B_{m}(t)= \{ 0, B \}.
\end{eqnarray}

To minimize the computational resource of searching, we restrict our control on the first and last 3 spins of the chain, i.e.~only the magnetic fields on a total of 6 spins are allowed to switch. This is because the transfer of the state in the middle of the chain is unlikely to be disturbed \cite{Murphy.10}. Therefore, we are  dealing with 6 control parameters instead of the two in the case of  \cite{Caneva.09,Murphy.10}. Furthermore, the Krotov algorithm is based on gradient descendent and it is not efficient for our discrete problem at hand. We therefore need to seek a different approach.

\subsubsection{Markov Decision Process (MDP)} \label{sec:deep}

In the following, we show how the state transfer problem with time-dependent Hamiltonian can be converted  into an MDP. 
An MDP contains a state space $\mathcal{S}$, an actions space $\mathcal{A}$ and a reward function $r : \mathcal{S}\times \mathcal{A}\rightarrow \mathbb{R}$. Consider an agent at state $s\in \mathcal{S}$ and takes action $a\in\mathcal{A}$, its resulting state $s'\in\mathcal{S}$ is then given by the transition probability $p(s'|s,a)$. If the transition process is deterministic, $s'$ can be written as a function of $s$ and $a$:
\begin{equation}
s'=s'(s,a).\label{eq:st}
\end{equation}
Then, given $s$, $a$ and $s'$, the reward can generally be written as:    
\begin{equation}
r=r(s,a,s').\label{eq:genr}
\end{equation}
In other words, the resulting state and the immediate reward are uniquely determined by the current state of the agent and the action it takes, so the process is Markovian. Should the markovianity be violated, the agent would not receive enough information to make decisions of the action it would take, and RL would likely break down \cite{Sutton,Whitehead.95}.

At each time step $t$, the agent chooses an action $a_{t}\in \mathcal{A}$ according to its current state $s_t\in\mathcal{S}$ and transition policy $\pi(a|s)$, which is essentially a list of probabilities that the agent takes action $a$ when it is at the state $s$. Its aim is to maximize the total discounted reward: 
\begin{equation}
R_t=\sum_{i=t} \gamma^{i-t} r_{i},
\end{equation}
where $r_i$ is the immediate reward at time $i$, and $\gamma$ is the discount factor. The discount factor indicates that as time elapses, the agent will receive less and less immediate reward. So generally, it would favor a policy with  a shorter total time. 

(a) \textit{Agent's state}. The design of the agent's states should ensure the markovianity. In our problem, all information is stored in the wave function, so we need an $s_{t}$ to represent $|\psi_{t}\rangle$. To facilitate the input to the neural network (to be discussed later), $s_{t}$ is defined as a real vector:
\begin{equation}
\begin{split}
s_{t}=f(|\psi_{t}\rangle)=&[\text{Re}(|\psi_{t}(1)\rangle),\ldots\text{Re}(|\psi_{t}(N)\rangle),\\
&\text{Im}(|\psi_{t}(1)\rangle),\ldots\text{Im}(|\psi_{t}(N)\rangle) ]^{T},\label{eq:state}
\end{split}
\end{equation}
where $|\psi_{t}(n)\rangle$ is the complex amplitude of wave function $|\psi_{t}\rangle$ at site $n$.

(b) \textit{Agent's action}. In our problem, the evolution operator $U_{t,t+1}$ is determined by the configuration of the external magnetic field, which consequently determines the action our agent shall choose. As explained above, we either control the first or the last three spins in the course of the transfer, so we have a total of 16 allowed actions ($0\leq a_{t}\leq15$) defined as follows. When $0\leq a_{t}<8$, the first three spins satisfy:
\begin{equation}
a_{t}=\frac{1}{B}\left[2^{0}B_{1}(t)+2^{1}B_{2}(t)+2^{2}B_{3}(t)\right],
\end{equation}
while $B_{m>3}(t)=0$. When $8\leq a_{t}<15$, the last three spins satisfy:
 \begin{equation}
 a_t-7=\frac{1}{B}\left[2^{0}B_{N-2}(t)+2^{1}B_{N-1}(t)+2^{2}B_{N}(t)\right],
 \end{equation}
  while $B_{m<N-2}(t)=0$. When $a_{t}=15$, $B_{m}(t)=B$ for all sites, meaning that the magnetic fields for all sites are turned on.
 Together with the definition of $s_{t}$ in Eq.~\eqref{eq:state} and the evolution $|\psi_{t+1}\rangle=U_{t,t+1}|\psi_{t}\rangle$, the transition function Eq.~\eqref{eq:st} can be specified. 
  
(c) \textit{Reward}.
Judiciously engineering a reward function is important for RL to run effectively \cite{Dewey.14}. While the reward is in general a function of the current state, action and the resulting state [cf.~Eq.~\eqref{eq:genr}], in our problem it is sufficient to consider the reward dependent on the current state only, $r=r(s)$. In a simplest method of giving rewards, called sparse reward \cite{Vevcerik.17}, the agent will not receive any reward unless the fidelity is above certain threshold. Nevertheless, learning based on this method is inefficient because the agent may spend a great amount of effort without gaining any reward until it hits the correct path by coincidence. 

This strategy is modified as follows. At each time step, a relatively small reward is given if the fidelity is bellow the threshold. However, if the fidelity is above the threshold, the agent will receive a much higher reward, and the current episode is terminated. We found that the following three-piece reward function worked well: 
\begin{align}
r= \left\{
\begin{array}{rcl}
10F(t)     &    & F(t)\leq 0.8 \\
\frac{\displaystyle 100}{\displaystyle (1+\exp\left[10(1-\xi-F(t))\right]}   &  & 0.8\leq F(t)\leq 1-\xi\\
2500     &  &  F(t)> 1-\xi\\    
\end{array} \right. \label{eq:R}
\end{align}
where $\xi$ is the infidelity threshold.

 In practical implementations, all functions and evolutions of the agent are encapsulated in a \emph{class} of our Python program. These algorithms are classical ones in the sense that they are all realized on a classical computer without any quantum-mechanical process involved.

\subsubsection{Q-learning} \label{sec:deep}

Q-learning is one of the most studied RL algorithms. It solves the MDP problem by maximizing the action-value function.
For a particular policy $\pi(a|s)$, the action-value function $Q^{\pi}(s,a)$ is defined as the total reward under policy $\pi$ with initial state $s$ and action $a$: 
\begin{equation}
Q^{\pi}(s,a)=\mathbb{E}[R_{t}|s_{t}=s,a_{t}=a,\pi].
\end{equation}
The optimal action-value function, $Q^{*}(s,a)= \max_\pi Q^{\pi}(s,a)$, satisfies the so-called Bellman equation \cite{Sutton}:
\begin{equation}
Q^{*}(s,a)= r+\gamma \max_{a^{'}}Q^{*}(s',a'),\label{eq:Bell}
\end{equation}
where $a'$ is the next action. $Q^{*}(s,a)$ therefore encapsulates two key pieces of information for the learning: the maximal total reward that can be obtained, and the best action the agent can take (i.e., the action $a$ which maximizes $Q^{*}(s,a)$) when it is at state $s$.

 In conventional Q-learning, $Q(s,a)$ is represented by a table called the $Q$-table, each element of which represents the $Q$-value for a pair $(s,a)$. In other words, the $Q$-table keeps the one-to-one correspondences between  $(s,a)$ and $Q(s,a)$. However,  $s$ represents wave functions and has a high dimensionality, so it is impossible to write down all possible values of $s_t$ [cf.~Eq.~\eqref{eq:state}] in the $Q$-table. For many physically relevant problems including the one at hand, the $Q$-table would be too large to maintain, which necessitates an alternative method, called the Q-network, to be introduced in the next subsection.
 
\begin{algorithm} [!htbp]
\caption{Deep RL for dynamical control} 
\label{alg2} 
\begin{algorithmic}

\STATE Initialize memory $R$ to empty

Randomly initialize the evaluation network $\theta$

Initialize the target network $\theta^{-}$ by: $\theta^{-}\leftarrow\theta$

\FOR{episode$=0$, $M$} 
\STATE \quad Initialize $|\psi_{0}\rangle$ , $s_{0}=f(|\psi_{0}\rangle)$

\quad\textbf{for} $t=0$, $t_{\max}$, \textbf{do}

\quad \quad With probability $\epsilon$ select a random action $a_{t}$,

\quad \quad otherwise  $a_{t}=\text{argmax}_{a}Q(s_{t},a;\theta)$

\quad \quad Execute $a_{t}$ and observe the reward $r_{t}$, and the next state $s_{t+1}$

\quad \quad Store experience $e_{t}=(s_{t},a_{t},r_{t},s_{t+1})$ in $R$

\quad \quad \textbf{if} $t$ is divisible by $t'$

\quad\quad \quad Sample minibatch of experiences  $e_{i}$

\quad\quad \quad Set $y_{i}=r_{i}+\gamma \max_{a'}\hat{Q}(s_{i+1},a';\theta^{-})$

\quad\quad \quad Update $\theta$ by minimizing $L=\left[y_{i}-Q(s_{i},a_{i};\theta)\right]^{2}$

\quad \quad \textbf{end if}

\quad \quad Every $C$ times of learning, set $\hat{Q}\leftarrow Q$

\quad \quad \textbf{Break} if $1-F(t)<\xi$

\quad  \textbf{end for} 
\ENDFOR 

\end{algorithmic} 
\end{algorithm}

 \subsubsection{Deep Reinforcement learning (deep RL)}
 Dramatic improvements can be made when the $Q$-table is replaced by a deep neural network (named the deep Q-network). Instead of keeping a table that records the one-to-one correspondence between $(s,a)$ and $Q(s,a)$, this correspondence is maintained by a neural network, and $s$ becomes its input while $a$ is the output, and $Q(s,a)$ the nonlinear relation between them. The introduction of a neural network empowers us to treat the large state space as the network can first be trained using a limited set of data and is then capable to predict $Q(s,a)$ for a larger set of inputs $s$. The Q-learning that uses the Q-network is called the ``deep Q-learning'', and the combination of deep neural networks and RL is termed as ``deep RL''. 

Our deep Q-learning algorithm is based on the Bellman function Eq.~\eqref{eq:Bell}. An evaluation network $\theta$ is used to represent $Q^{*}(s,a)$, and a target network $\theta^{-}$ is for $Q^{*}(s',a')$. The state vector $s$ ($s'$) serves as the input of $\theta$ ($\theta^{-}$), and each neuron at the output layer corresponds to a particular choice of action $a$. The outputs of these neurons $Q(s,a;\theta)$ and $\hat{Q}(s',a';\theta^{-})$  serve as estimates of the $Q$-value provided by agent. Since the optimal policy satisfies Eq.~\eqref{eq:Bell}, one can update the evaluation network $\theta$ by minimizing the loss function: 
\begin{equation}
L= \left[y_{i}-Q(s,a;\theta) \right]^2, \label{eq:L}
\end{equation}
where
\begin{equation}
y_{i}=r_{i} +\gamma \max_{a'} \hat{Q}(s_{i+1},a';\theta^{-}).\label{eq:y}
\end{equation}

The complete algorithm of our deep Q-learning is shown in Algorithm~\ref{alg2}. There are generally two tasks for the agent: execution and learning. For execution (see Fig.~\ref{fig:sketch}(a)), the agent chooses the action $a_{t}$ according to the $\epsilon$-greedy policy: with probability $\epsilon$, we choose a random action; otherwise, we choose the action with highest action-value function: $a_{t}=\text{argmax}_{a}Q(s_{t},a;\theta)$. The configuration of the magnetic fields and $U_{t,t+1}$ is then determined. Under $U_{t,t+1}$, the system evolves to $|\psi_{t+1}\rangle$ and $s_{t+1}=f(|\psi_{t+1}\rangle)$ is exported, which is then used by the agent for the next time step of decision making. The interpreter then generates an immediate reward $r_{t}$ according to Eq.~\eqref{eq:R}. The experience $e_{t}=(s_{t},a_{t},r_{t},s_{t+1})$ as a whole is stored in its memory $R$. The agent terminates its current episode if the infidelity is below $\xi$, and begins a new episode with initial state $s_{0}=f(|\psi_{0}\rangle)$.

During the course of learning (see Fig.~\ref{fig:sketch}(b)), the agent updates the deep neural network $\theta$ to make increasingly better decisions. At each learning step, the agent randomly retrieves a minibatch of experiences $e_{i}=(s_{i},a_{i},r_{i},s_{i+1})$ from memory $R$ according to a rule called ``prioritize experience replay'' (see \cite{Schaul.15} for details).
 $s_{i+1}$ is fed to the target network $\theta^{-}$ to calculate the target value $y_{i}$ according to Eq.~\eqref{eq:y}. At the same time, $s_{i}$ is sent to the evaluation network for calculation of $Q(s_{i},a_{i};\theta)$. Then, the evaluation network $\theta$ is updated through the 
 back-propagation  by minimizing $L$ as defined in Eq.~\eqref{eq:L}. In contrast, the target network $\theta^{-}$ is not updated in every step; instead, it is copied from the evaluation network every $C$ steps. The details of all hyper-parameters for Algorithm~\ref{alg2}  can be found in Table~\ref{table}. 

In the literature, numerical solutions to the problem of quantum state transfer typically involve gradient-based optimal control schemes, such as GRAPE \cite{Bukov.17}, quasi-Newton \cite{Wang.10spin,heule.10}, and iterative algorithms for inverse eigenvalue problems \cite{kay.06} (apart from the Krotov method to be discussed in Sec.~\ref{sec:resdy}). Here, we briefly comment on the similarity and difference between them and deep RL. On one hand, they all require little a priori knowledge of the problem: the value of a solution is measured by a cost function in the gradient based methods, while in deep RL it is represented by a reward function. On the other hand, there are fundamental differences. Firstly, Deep RL relies on a well-trained, highly non-linear deep neural network, which, to certain extent, is able to recognize and represent complex relations between its inputs and outputs. The neural network makes its prediction according to its own knowledge of the problem based on its vastly high-dimensional, nonlinear Q-network, which is constructed through extensive training. On the contrary, the gradient algorithm itself keeps no knowledge of the overall problem except the cost function and its gradient with respect to the parameters concerned. It is therefore believed that a judiciously trained deep neural network could be more powerful.  Secondly,  gradient-based algorithms typically requires the problem being continuous, namely the gradients of the cost function must be well behaved. Nevertheless, deep RL can in principle solve discrete or continuous problems. For example, in Sec.~\ref{sec:resdy} we consider a situation where the control field can only attain two values (``on'' and ``off''), and conventional gradient-based algorithms are not readily applicable.

\begin{figure}
\includegraphics[width=0.9\columnwidth]{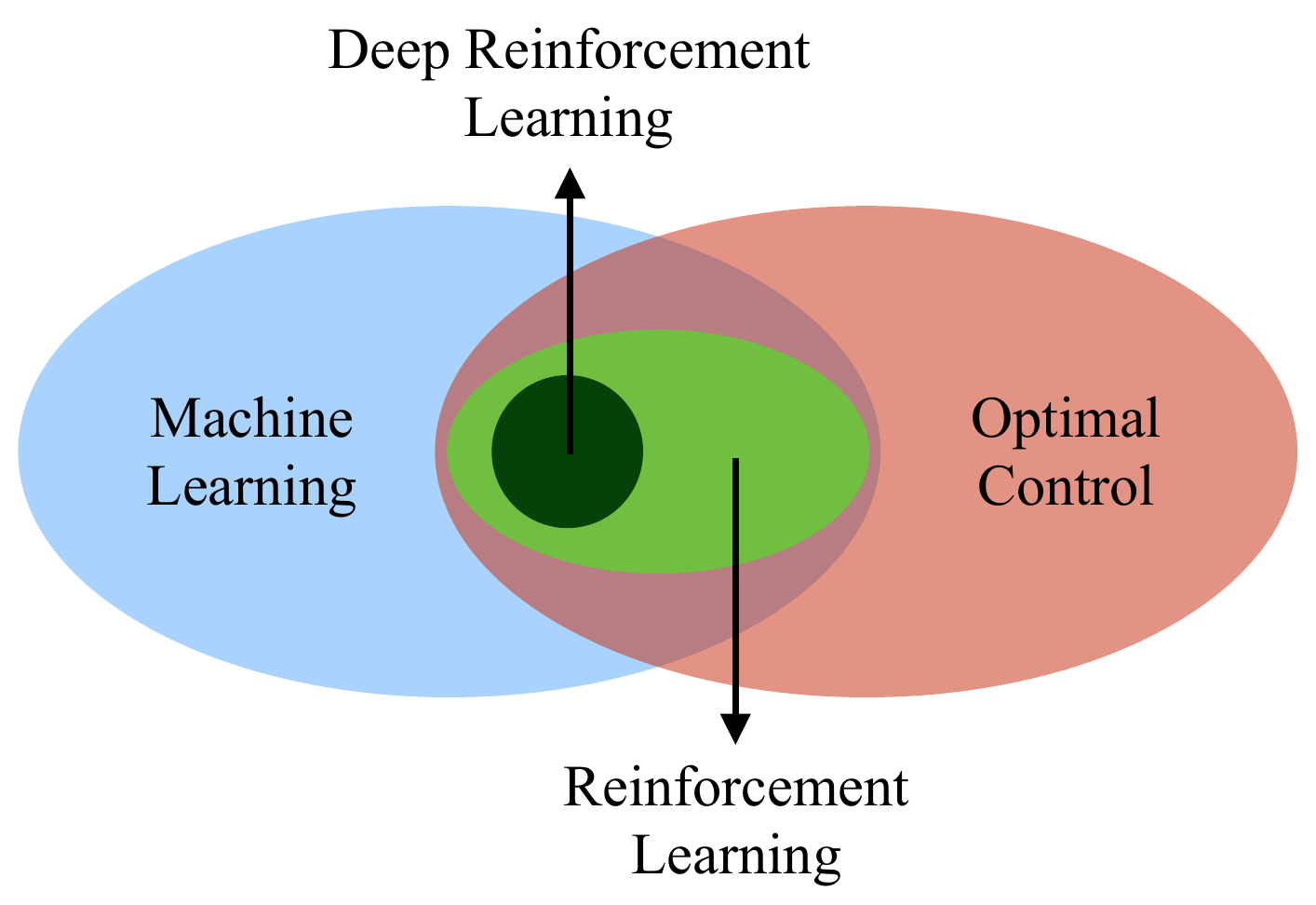}
\caption{Venn diagram showing the relationships between optimal control, machine learning, RL and deep RL.}
\label{fig:relation}
\end{figure}

Before we end this chapter we briefly discuss the relationship between optimal control, machine learning (used interchangeably with ``artificial intelligence'' in most places of this paper), RL and deep RL, which is shown as a Venn diagram in Fig.~\ref{fig:relation}. As discussed above, many methods can be used for optimal control, with machine learning being one candidate. RL is one type of machine learning technique, which we have employed to the optimal control problem in this work. Deep RL is a sophisticated version of RL that (mostly) uses Q-networks instead of Q-tables.

\begin{table}
\caption{\label{table}List of hyper-parameters for deep RL}
\begin{ruledtabular}
\begin{tabular}{ccl}
 Hyper-parameter &Notation in & \mbox{value} \\
   &  main text&  \\
\hline
Minibatch size& &32\\
Replay memory size&& 40000  \\
Learning rate in back propagation && 0.01\\
$\theta^{-}$ replace period &$C$& 200\\
Reward decay &$\gamma$& 0.95\\
Number of hidden layer & & 2\\
Neuron number per hidden layer & &120\\
$\epsilon$-greedy rate  &$\epsilon$& \footnote{ $\epsilon$ is initialized to $1$, and after each step of learning, we reset $\epsilon \leftarrow \epsilon -0.0001$ until $\epsilon=0.01$.}\\ 
Learning period  &$t'$&5\\
Maximum steps &$t_{\max}$&\footnote{From $N=5$ to $N=11$, $t_{\max}$=24, 28, 32, 38, 44, 50, 50 respectively.}\\
Total episode  &M&50000\\

Time step&$dt$ &0.15\footnote{ $dt$ at the final steps of each episode are set differently. It is chosen within $dt\in[0.015, 0.3]$ that give us the highest transfer fidelities.}\\
Coupling strength &$J$& 1\\
Magnetic field strength& $B$&100\\
\end{tabular}
\end{ruledtabular}
\end{table}

\begin{figure}
\includegraphics[scale=0.35]{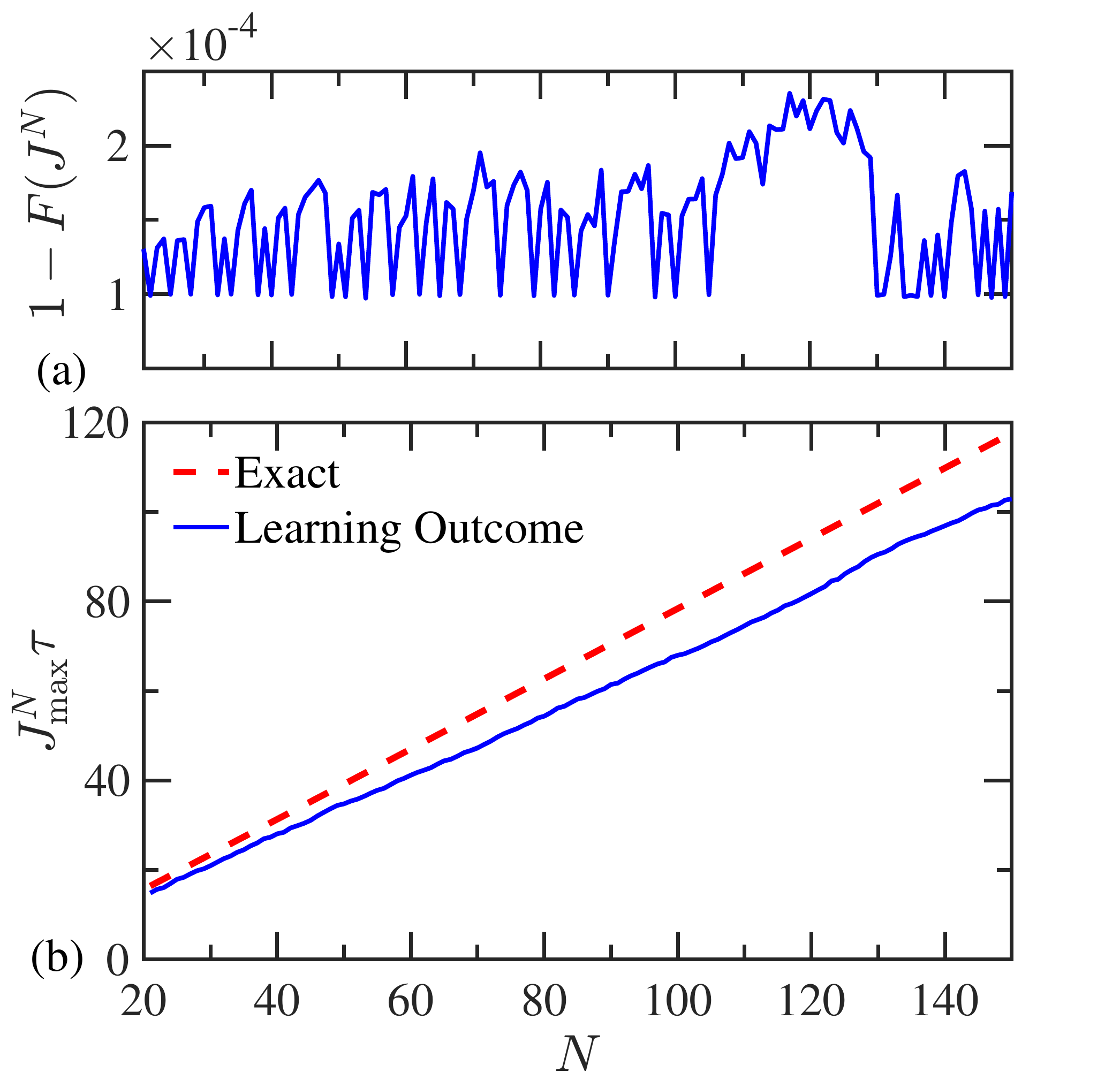}
\caption{\label{fig:spind}  (a) Infidelity of our learning outcome as a function of $N$, the size of the spin chain. Note that all infidelity values are below  $5\times10^{-4}$.
(b) Comparison of the efficiency between the exact result \cite{Christandl.04} and our learning outcome. Red dash line: efficiency for the exact result \cite{Christandl.04} while keeping the infidelity below $5\times10^{-4}$. Blue solid line: efficiency for our learning outcome.}
\end{figure}

\begin{figure}
\includegraphics[scale=0.4]{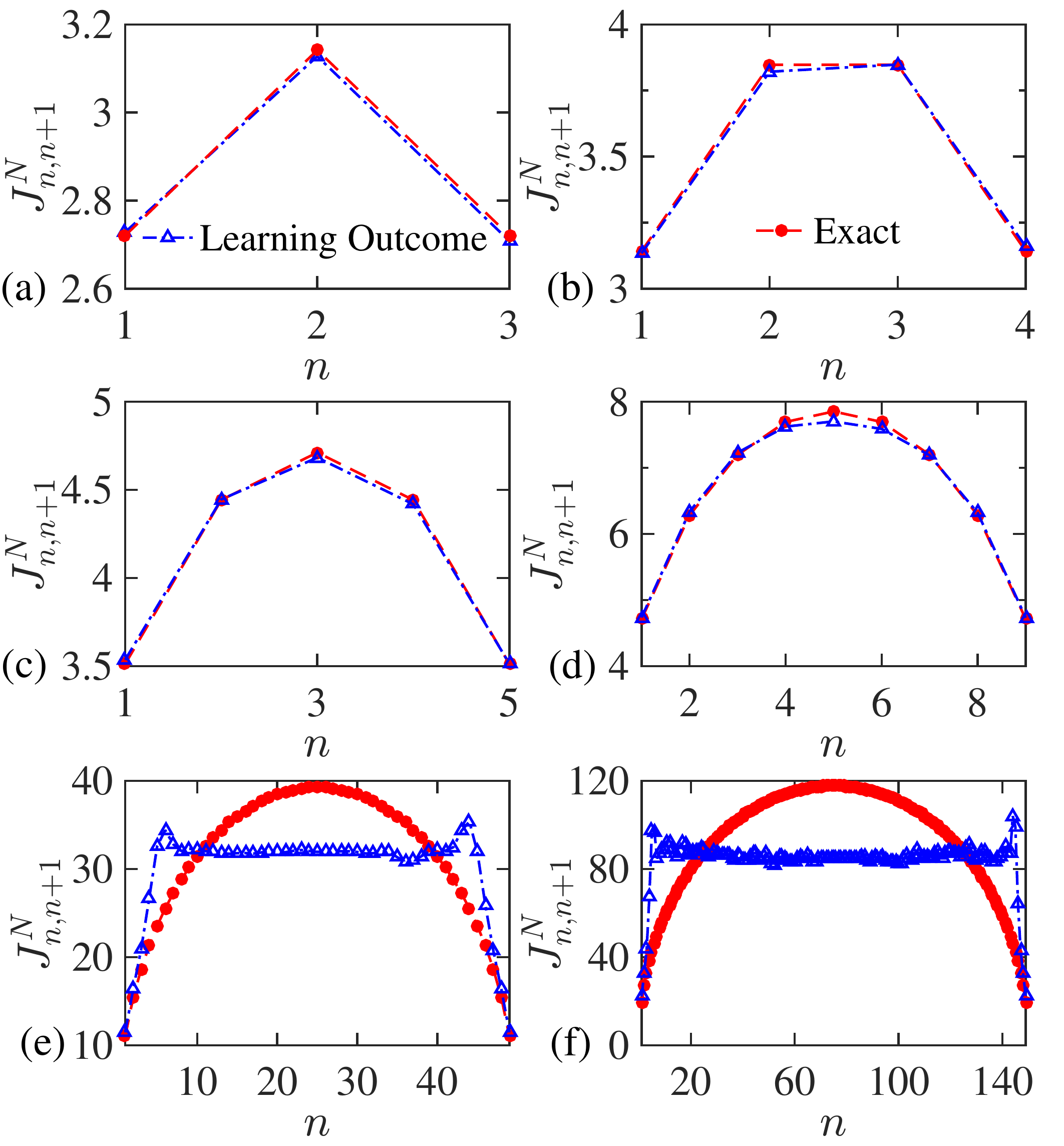}
\caption{\label{fig:shape} Calculated coupling strengths for spin transfer in the time-independent case. The coupling strengths $J^N_{n,n+1}$ are shown site-wise for (a) $N=4$, (b) $N=5$, (c) $N=6$, (d) $N=10$, (e) $N=50$, and (f) $N=150$. Blue dash-dotted line: the learning outcome from  Algorithm~\ref{alg1}. Red dashed line: the coupling schemes in \cite{Christandl.04}.}
\end{figure}

\begin{figure} [!htbp]
\includegraphics[scale=0.4]{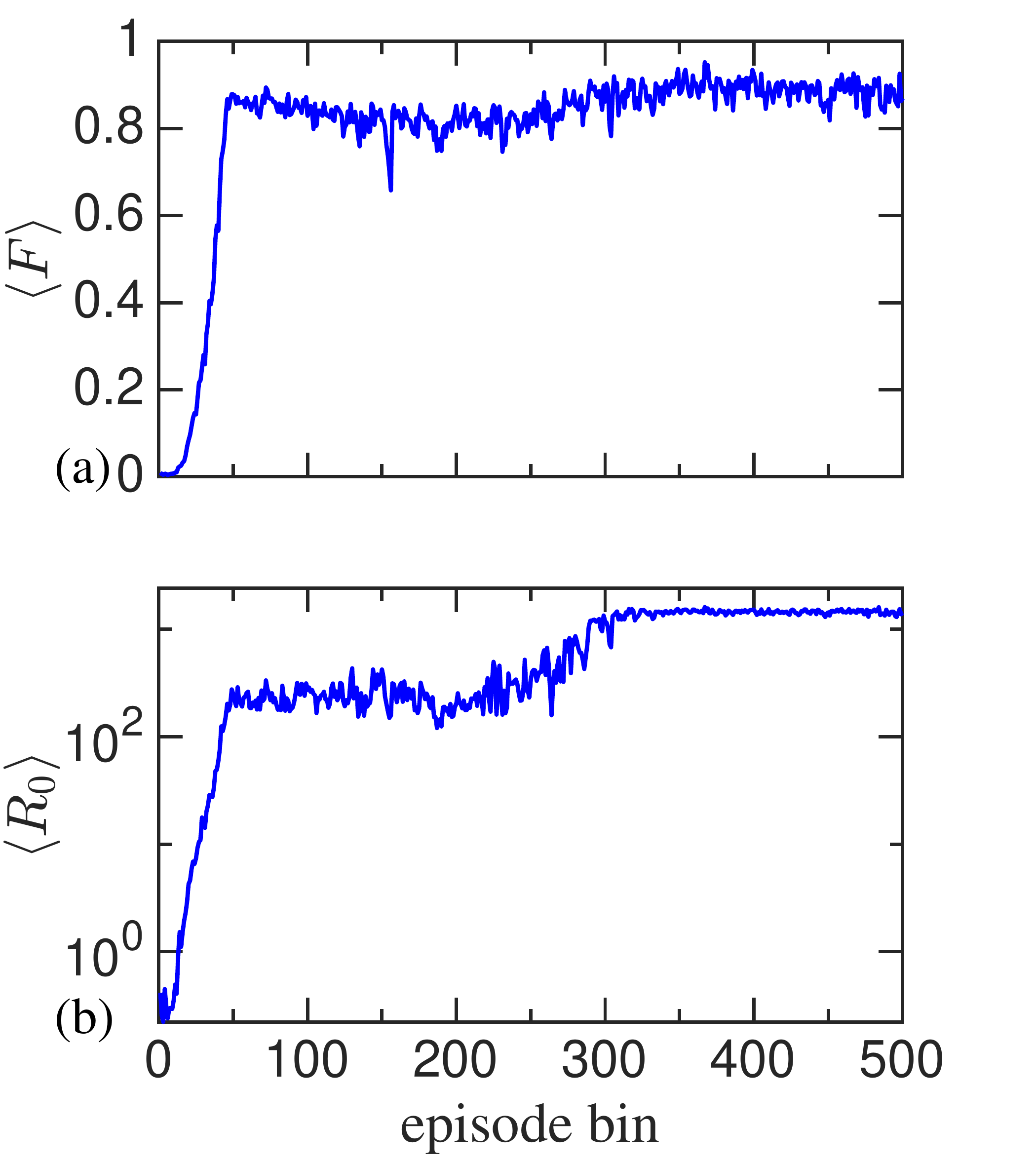}
\caption{ (a) The averaged final fidelity as a function of the number of episode bins in the time-dependent case. (b) Average total reward v.s. number of episode bins in the time-dependent case. $N=9$. For details on how we bin our data and the hyper-parameters, see Sec.~\ref{sec:resdy} and Table~\ref{table}.}
\label{fig:curve}
\end{figure}

\section{Results}
\label{sec:res}

Our algorithms are implemented with Python 2.7, and have been run on a 8-core 3.60GHz CPU with 7.7 GB memory. The runtime for algorithm~\ref{alg1} with $N_{\max}=150$ is about few hours; and for algorithm~\ref{alg2} of single experiment it is within an hour.

\subsection{Time-independent case}
\label{sec:resind}
We have applied Algorithm~\ref{alg1} for $N_{\max}=150$. The perturbation strength, learning rate, relaxation parameter and precision are set as $\beta = 1$, $\alpha=0.05$, $\gamma = 10^{-6}$, and $\xi=10^{-4}$ respectively. As shown in Fig.~\ref{fig:spind}(a), the infidelity of learning outcomes can always be maintained below $5\times 10^{-4}$, and has no obvious increase as the chain becomes longer. 

In Fig.~\ref{fig:spind}(b), we compare the transfer efficiency ($J^{N}_{\max}\tau$) of learning outcomes and the exact scheme in \cite{Christandl.04}, which has proven to be the QSL for perfect state transfer with time-independent Hamiltonian \cite{Yung.06}. Here $\tau$ for the exact scheme is defined as the minimal transfer time while maintaining infidelity $<5\times 10^{-4}$. The efficiencies of our learning outcomes are always better than the exact scheme (smaller efficiency is better because it means shorter time when $J_{\rm max}$ is fixed.). For $N=150$,  the efficiency of learning result is $0.88$ times that of \cite{Christandl.04}. In other words, \emph{the transfer speed can be enhanced by more than $10\%$ by sacrificing only $\sim10^{-4}$ of infidelity}. Furthermore, the slope of the learning outcome ($0.69$) is smaller than the exact scheme ($0.78$), indicating a lower transfer time per site during the process. 

\begin{figure} [t]
\includegraphics[scale=0.4]{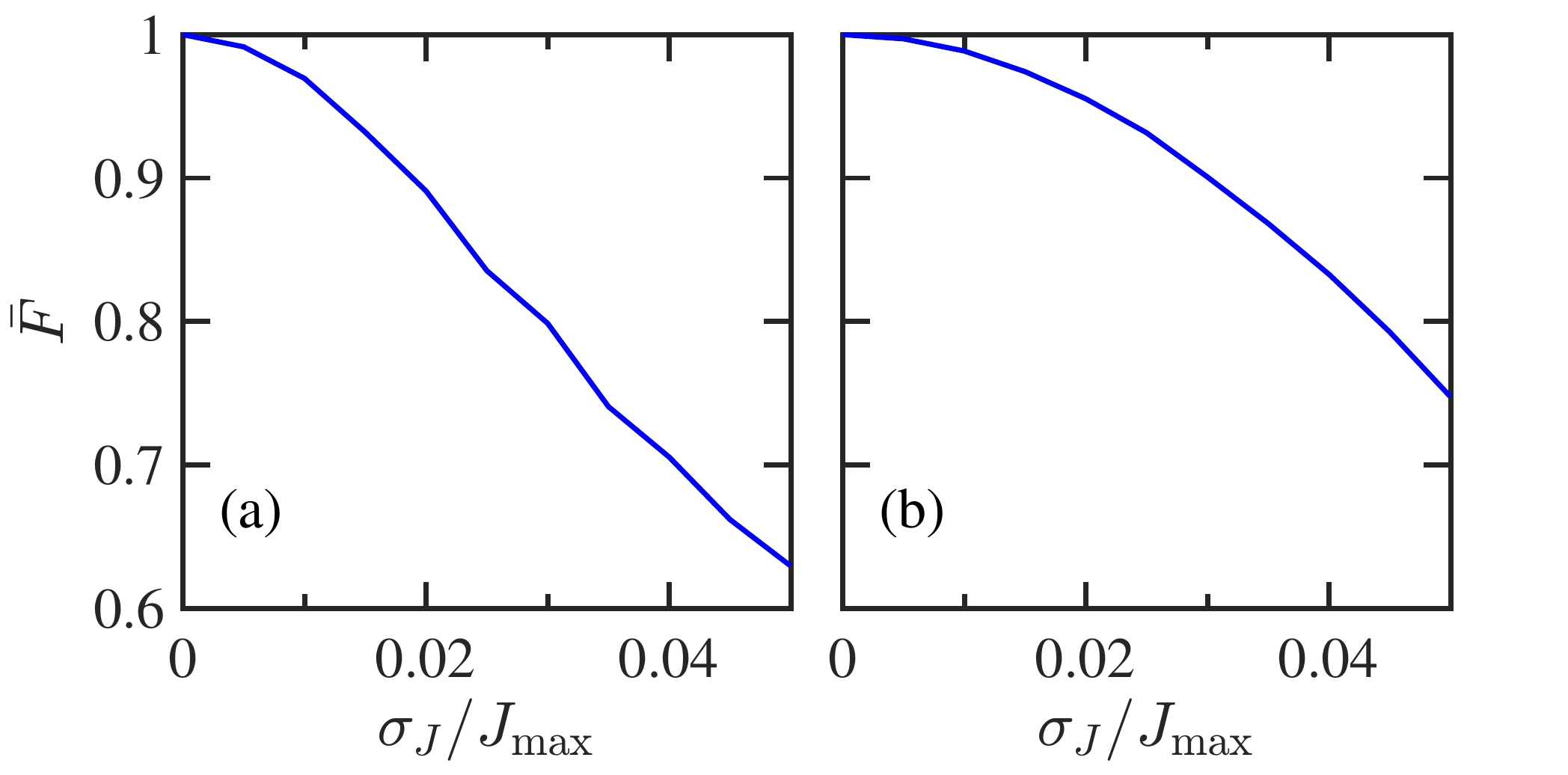}
\caption{Average fidelity of the spin transfer v.s. the amplitude of the fluctuation of the coupling strengths in Eq.~\eqref{eq:tindepHammat}. (a) Fluctuations are uniform, i.e. all couplings have the same deviation. (b) Coupling strengths fluctuate individually.
}
\label{fig:stability1}
\end{figure}

The comparison of the coupling strength between our learning outcomes and the exact scheme in \cite{Christandl.04} are shown in Fig.~\ref{fig:shape}. For $N<10$, there are no obvious differences between the learning outcomes and the exact ones. However, the curves showing coupling strengths begin to deviate for $N\geq10$. Instead of a smooth parabolic shape, the coupling strength of the learning outcomes abruptly increases to a high value, then remains almost constant in the middle of the chain, and decrease abruptly at the end. We believe that it is the flatness of the curve in the middle of the chain that is responsible for the  speeding up. For most of the time in the middle, the coupling strengths are kept  close to its maximal value. Since stronger coupling strength implies higher transfer speed, the total spin transfer process will be completed faster, although the transfer is no longer 100\% perfect. We note that the similar pattern of the coupling strengths has been observed in (mostly analytical) studies of spin transfer and long-range entanglement on spin chains \cite{wojcik.05,campos.07,bayat.10}. In \cite{wojcik.05,campos.07} it has been found that near-perfect spin transfer can occur on a chain with the coupling strengths at the two ends close to zero while being uniform otherwise. It is interesting to note that our learning algorithm produces similar results without any a priori knowledge except the Hamiltonian per se. Moreover, the optimal coupling strengths on the ends that we have found are relatively small but not close to zero, suggesting that the actual transfer in our scheme may be faster. The optimal end coupling strengths also appear in closely related contexts such as entanglement gates between distant qubits realized on cold atom chains \cite{Banchi.11}.  

We now discuss the stability of our learning outcome against fluctuations in the coupling strengths. This fluctuation is introduced in  the model [Eq.~\eqref{eq:tindepHammat}] by replacing the coupling strength $J_{i,i+1}^{N}\rightarrow J_{i,i+1}^{N} +\delta J_{i,i+1}^{N}$ on top of the training outcome from the ideal Hamiltonian. The results are shown in Fig.~\ref{fig:stability1}, where the $x$-axes are the amplitude of the error (to be discussed below), while the $y$-axes are the fidelity of the transfer that is averaged over 1000 runs. Fig.~\ref{fig:stability1}(a) shows the case where the fluctuations are uniform, i.e. $\delta J_{i,i+1}^{N}=\delta J$ with $\delta J$ drawn from a normal distribution $\mathcal{N}\left(0,\sigma_{J}^{2}\right)$. We see that the averaged fidelity remains abot 95\% for $\sigma_J/J_{\rm max}<1\%$, but drops to about 63\% as $\sigma_J/J_{\rm max}$ is increased to $5\%$. In Fig.~\ref{fig:stability1}(b), the coupling strengths are allowed to fluctuate individually $\left[\delta J_{i,i+1}^{N}\sim\mathcal{N}\left(0,\sigma_{J}^{2}\right)\right]$. In this case, the average fidelity drops slower than the previous one, being about 95\% for $\sigma_J/J_{\rm max}\approx2\%$ and drops to 75\% at $\sigma_J/J_{\rm max}\approx5\%$. These results suggest that our transfer scheme is reasonably robust to noises provided that the noise is below 2\%.

\begin{figure*}
\includegraphics[scale=0.85]{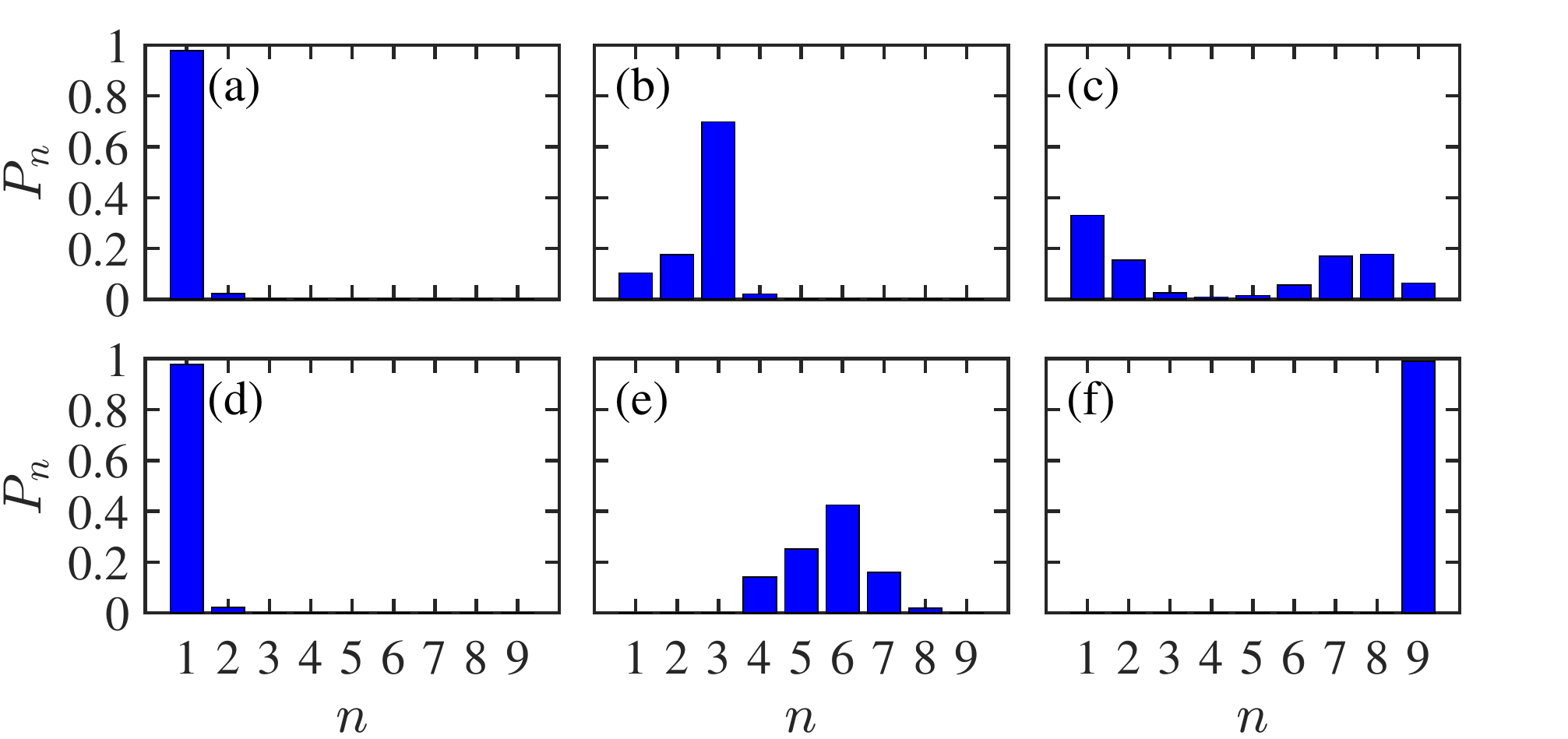}
\caption{ The probability distribution in the time-dependent case. (a)-(c) The data is drawn at the 1000th episode of learning and the transfer fails. (a) the first step of the transfer. (b) the 22nd out of a total of 44 steps. (c) the 44th step, i.e.~the final step of the transfer.
(d)-(f) The data is drawn after the agent has been well-trained. The transfer succeeds. (a) the first step of the transfer. (b) the 20th out of a total of 40 steps. (c) the 40th step, i.e.~the final step of the transfer. }
\label{fig:packet}
\end{figure*}

\subsection{Time-dependent case}
\label{sec:resdy}
To make the learning process smooth, we need to be careful in setting parameters to the deep RL. For example, if the infidelity threshold $\xi$ in Algorithm \ref{alg2} is too high, the algorithm will break early, and the resulting fidelity of the learning outcome is far from ideal. On the contrary, if $\xi$ is too small for the agent to reach, the reward gained remains low and the learning is inefficient. We have found that $\xi=0.05$ makes a reasonable compromise. 

We have performed Algorithm \ref{alg2} for different lengths ($N$) of spin chains. From $N=5$  to $N=11$, the highest fidelities we obtained are $F=0.9985, 0.9961, 0.9934, 0.9907, 0.9903, 0.9838$ and $0.9664$ respectively.
To display the effect of learning, we trace both the final fidelity ($F$) and the total discounted reward ($R_{0}$) of each episode. We bin the data for every 100 episodes, and calculate the mean value of the final fidelity ($\langle F\rangle$) and the total discounted reward ($\langle R_{0}\rangle$) within each episode bins. Examples of the learning curves of $\langle F\rangle$ and $\langle R_{0}\rangle$ for a nine-spin chain are shown in Fig.~\ref{fig:curve}. It can be seen clearly that as the learning episodes increase, both $\langle F\rangle$ and $\langle R_{0}\rangle$ increase, which more or less saturate at high values. We therefore conclude that the agent has successfully learned an efficient policy for our problem.

\begin{figure}
\includegraphics[scale=0.45]{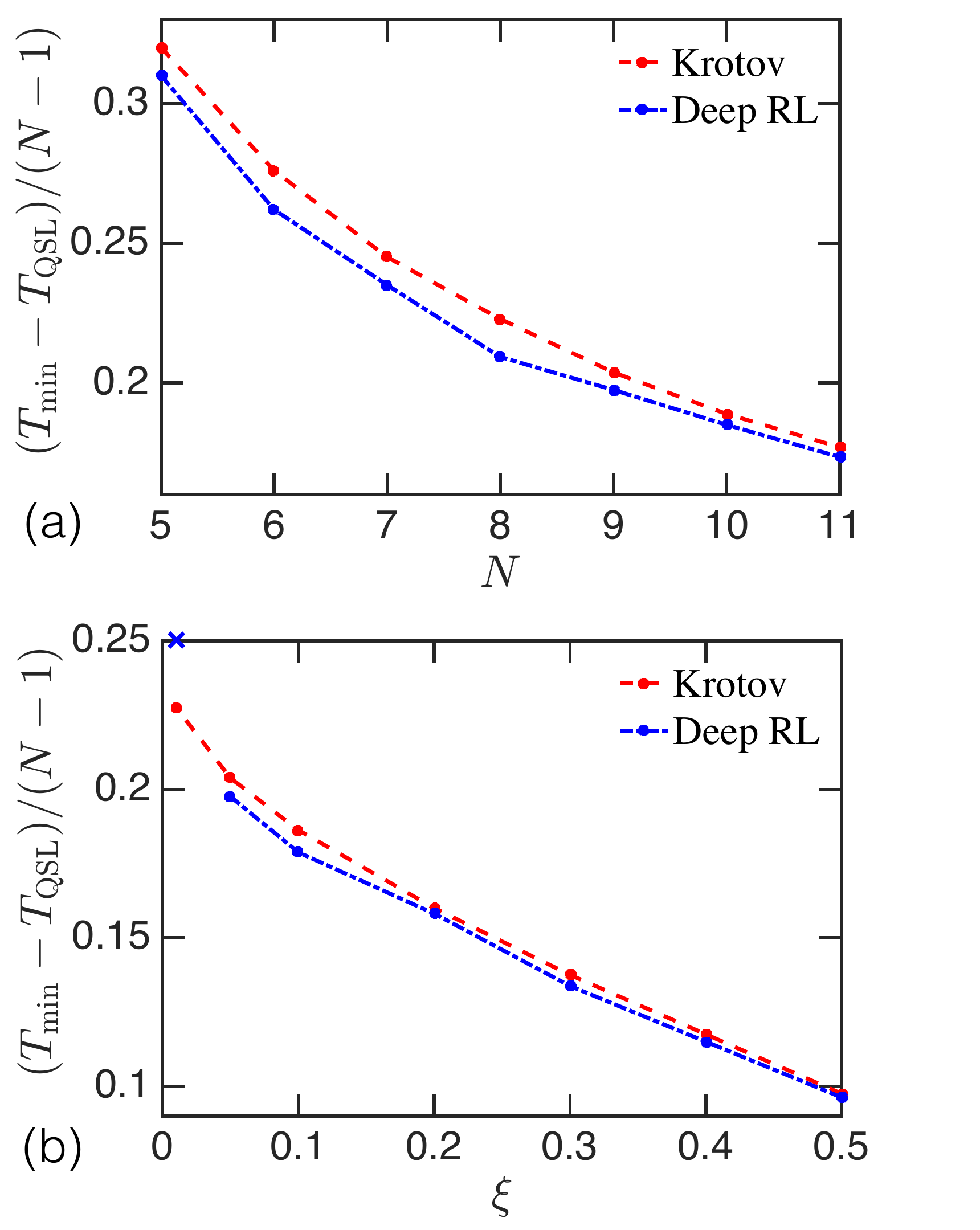}
\caption{(a) Minimum transfer time maintaining infidelity$<5\%$ for different values of $N$. (b) Minimum transfer time for $N=9$ and maintaining infidelity$<\xi$ for different values of $\xi$. Red dash line: with method in Ref.~\cite{Caneva.09,Murphy.10}; blue dash-dot line: with deep RL. Blue cross is the transfer time for episode with highest fidelity with deep RL.}
\label{fig:spdy}
\end{figure}

\begin{figure}
\includegraphics[scale=0.4]{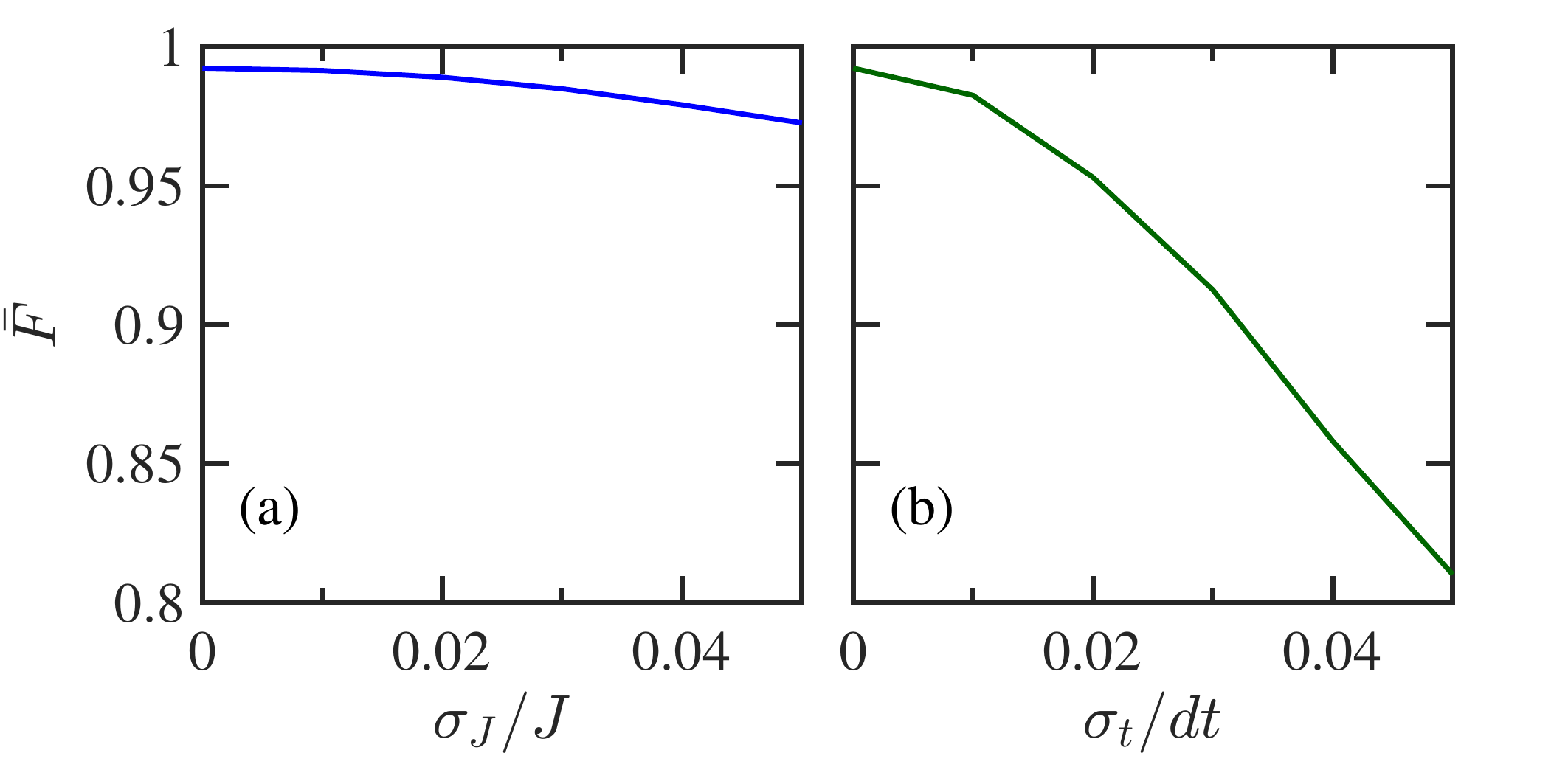}
\caption{Average fidelities of the spin transfer v.s. the amplitude of the fluctuations of (a) the coupling strengths and (b) the pulse (time-dependent magnetic field) turn on/off time. In (a), fluctuations on couplings are assumed to be uniform and are drawn from a normal distribution.}
\label{fig:stability2}
\end{figure}

To study the learning process in more detail, we investigate the status of the transfer at different learning stages.  In Fig.~\ref{fig:packet}, we compare the distribution of $P_{n}$, which is defined as $P_{n}=|\langle\psi_{t}(n)|\psi_{t}(n)\rangle|^{2}$, between the early learning stage ($\text{episode}=1000$) and an episode after the agent is well-trained. 
The first row of Fig.~\ref{fig:packet} shows sequentially representative results during the initial, intermediate and final stage of the transfer when the learning is not successful.  $P_{n}$ is distributed randomly and the transfer failed in the end. The second row of Fig.~\ref{fig:packet} shows the results when the  agent is well-trained. It is clear that the wave packet passes  the middle sites during the transfer, and the state has been successfully transferred to the other end of the chain eventually.

As mentioned above, the agent favors policies with shorter time steps because of the reward discount. In order to reveal the effectiveness of deep RL, we employ the method of  \cite{Caneva.09,Murphy.10} and calculate the implied transfer time. The shortest transfer time $T_{\min}$ under infidelity $5\%$ for both deep RL and those implied by \cite{Caneva.09,Murphy.10} are compared for spin chains with $5\le N\le11$. The results are shown in Fig.~\ref{fig:spdy}(a). Remarkably, the transfer scheme provided by our agent is always closer to  $T_{\rm QSL}$, namely, faster than that of \cite{Caneva.09,Murphy.10}. This fact holds even when the infidelity threshold $\xi$ is varied. In Fig.~\ref{fig:spdy}(b), we show the comparison of $T_{\min}$ versus the value of infidelity threshold $\xi$ for a nine-spin chain.
When $\xi$ is higher than 0.2, the reward function defined in Eq.~\eqref{eq:R} becomes inefficient. So we revise the reward function as:
\begin{eqnarray}
r= \left\{
\begin{array}{rcl}
10F(t)     &    & F(t)\leq 1-\xi \\
2500     &  &  F(t)> 1-\xi\\    
\end{array} \right. \label{eq:r}
\end{eqnarray}
The learning results in Fig.~\ref{fig:spdy}(b) always provide shorter transfer times from $\xi=0.5$ to $\xi=0.05$, although the difference of transfer time is larger for smaller $\xi$ when $\xi\gtrsim0.05$.  The blue cross corresponds to a scheme with highest fidelity ($F=0.9903$) the agent can find, which has a longer transfer time compared to the results implied by \cite{Caneva.09,Murphy.10}. 

We note that our method becomes relatively inefficient when the chain is long.  For $5\le N\le9$, our maximal fidelities obtained are greater than $0.99$, but this value drops to $0.9838$ for $N=10$ and $0.9664$ for $N=11$. Moreover, the difference in the transfer time between our results and those implied by \cite{Caneva.09,Murphy.10} is smaller when $N$ is close to 11, the longest chain we have studied. 

This loss of efficiency is expected, as we have confined our control to be on the first and last 3 spins only; for longer chains one must be capable to control more spins in order to complete the transfer efficiently. While it is always formidable to treat all spins at once, the problem can be simplified by only dealing with the few spins around the wave packet of the one being transferred. Nevertheless, this would still require a larger and possibly deeper Q-network, which is beyond the scope of this work. 

We discuss the stability of our transfer schemes against two types of fluctuations: the coupling strengths and the pulse (time-dependent magnetic field) turn on/off time. Fig.~\ref{fig:stability2} shows the results of running our transfer scheme in the presence of two types of fluctuations. The $y$-axes show the fidelity averaged over 1000 runs with different noise realizations. To facilitate the discussions, we choose one transfer scheme from the learning algorithm for which the fidelity is the highest.
 Fig.~\ref{fig:stability2}(a) shows the response of this transfer scheme to coupling fluctuations. We see that this transfer scheme is very robust to the coupling noises: the average fidelity remains above 95\% even the fluctuations in the coupling is as large as 5\%. As in Fig.~\ref{fig:stability1}(a), the fluctuations on the couplings are treated as equal, $J\rightarrow J+\delta J$, where $\delta J$ is drawn from $\mathcal{N}\left(0,\sigma_{J}^{2}\right)$. Nevertheless, this scheme is less robust to the fluctuations of the switching on/off time of the control fields, as can be seen from  Fig.~\ref{fig:stability2}(b). Here,  the delay or advance of the pulse switch on/off time is drawn from $\mathcal{N}\left(0,\sigma_t^{2}\right)$. While the average fidelity is about 95\% for $\sigma_t/dt\approx2\%$, it drops to approximately 81\% when $\sigma_t/dt\approx5\%$. These results show that the fluctuations in the pulse switch on/off time could be more damaging than those in the couplings.

\section{Conclusion and Outlook}
\label{sec:con}

To conclude, we have applied RL to the problem of quantum state transfer in a one-dimensional spin chain, and have successfully found transfer schemes with higher overall speed compared to previous results. For time-independent Hamiltonians, our RL algorithm updates the coupling strengths by minimizing a loss function taking into consideration both fidelity and efficiency. When the spin chain is long, the learning outcome gives coupling strengths that are different from one that has been previously proven to be the QSL for perfect transfer. We observed up to $10\%$ speed up while maintaining the infidelity below $5\times 10^{-4}$.  

For time-dependent Hamiltonians, we have demonstrated how this problem can be converted into a MDP. Deep Q-learning  is then used to update the transfer schemes by maximizing the estimated action-value function $Q(s,a;\theta)$. After proper training, the agent has successfully learned how a spin can be transferred efficiently while keeping infidelities to a minimum. The transfer speed we have obtained is closer to the maximal spin-wave speed than results implied by previous works based on Krotov method. 

We believe that the power of RL, especially deep RL, is yet to be unraveled in solving quantum physics problems. 
On one hand, although the Q-network may only be suitable to solve problems involving a discrete set of outputs so that the action space must be finite,  generalizations to problems involving continuous variables are not difficult using variations of deep RL \cite{Silver.14,Mnih.16}.
 On the other hand, the fact that deep neural networks are able to uncover the underlying structure of data \cite{Zhang.17(2)} suggests that deep RL may possess a similar potential that can ``learn'' from the data that the agent is fed. In other words, except for providing \emph{what} is the optimized outcome, it may also tell us the reason \emph{why} it is optimized. This can be potentially achieved by investigating the action-value function, $Q(s,a;\theta)$. While typically the action with the highest $Q$ is taken at each step, the $Q$ values that are close to the optimal one, as well as the corresponding location in the state space may hold important physical insights on the problem desired. 
Moreover, in cases involving noises, there have already been some learning algorithms proposed for robust control by simulating a large amount of parallel noise environments \cite{wu.17,niu.18}. Our results may provide insights on how these control schemes can be made more efficient. In a nutshell,
our results have demonstrated that RL is a capable tool to solve the problem of spin transfer, and we believe that further developments at the interface between artificial intelligence and physics shall grant us even more power to solve problems that have been previously deemed  difficult.

\section{acknowledgements}
This work is supported by the Research Grants Council of the Hong Kong Special Administrative Region, China (No.~CityU 21300116, CityU 11303617), the National Natural Science Foundation of China (No.~11604277, 11405093), the Guangdong Innovative and Entrepreneurial Research Team Program (No.~2016ZT06D348), Natural Science Foundation of Guangdong Province (2017B030308003), and the Science, Technology and Innovation Commission of Shenzhen Municipality (ZDSYS20170303165926217, JCYJ20170412152620376).

\end{document}